\documentclass[twocolumn,fleqn,usenatbib]{aastex631} 
\usepackage[T1]{fontenc}
\usepackage{physics}
\usepackage{tabularx}
\usepackage{graphicx}	
\usepackage{amsmath}	
\usepackage{amssymb}	
\usepackage{newtxtext,newtxmath}

\begin{document}

\title{Synchrotron break frequencies of mildly-to-highly relativistic outflows observed off-axis}

\author[0009-0003-0141-6171]{Gilad Sadeh}
\affiliation{Dept. of Particle Phys. \& Astrophys., Weizmann Institute of Science, Rehovot 76100, Israel}

\begin{abstract}
We consider the synchrotron spectrum produced by mildly-to-highly relativistic collisionless shocks. Simple analytic formulae are derived for the break frequencies (peak frequency, self-absorption frequency, synchrotron and inverse Compton cooling frequencies) of the emission produced by post-shock plasma elements propagating at an angle $\theta_e$ relative to the observer's line of sight. These formulae reproduce well the results of earlier exact analytic calculations
valid for ultra-relativistic shocks and also hold for 
$\gamma<10$ and for "off-axis" propagation (deviating from the ultra-relativistic results by approximately an order of magnitude). 
Our results will improve parameter estimation accuracy from future observations of synchrotron emission produced by collisionless shocks 
driven by the relativistic ejected material from compact objects mergers and jetted tidal disruption events. The improved accuracy for mildly relativistic velocities is essential since most events will be observed off-axis, with $\gamma<10$ outflows dominating the synchrotron emission (due to relativistic beaming). For GW170817, our results imply that (i) the Lorentz factor of the plasma emitting the observed radiation is bounded by $2.6<\gamma$ at $t\sim10$ days and by $\gamma<12$ at $t>16$ days, (ii) the interstellar medium (ISM) density, $n$, and the fraction of internal energy density held by magnetic fields, $\varepsilon_B$, are bounded by $n\cdot\varepsilon_B\lesssim 3\times10^{-7}$cm$^{-3}$. In future merger events in higher-density ISM, the peak and cooling frequencies may be identified in the radio and X-ray bands; consequently, $\gamma,n\cdot\varepsilon_B$ could be measured as opposed to the case of GW170817, where these frequencies are out of the observable range. 

\end{abstract}




\section{Introduction}
GW170817 provided a prolonged, rich, and unique set of thermal and non-thermal electromagnetic radiation observations.
During the first $\sim10$ days, observations of the optical (UV-IR) electromagnetic counterpart of GW170817 indicate ejecta of $\approx 0.05M_\odot$ expanding at $\beta>0.1$ and undergoing continuous heating by radioactive decays \citep{arcavi_optical_2017,nicholl_electromagnetic_2017,metzger_welcome_2017}. These findings are consistent with the characteristics of Kilonova (KN) emission \citep{li_transient_1998}.
The non-thermal emission observed between radio to X-ray is consistent with synchrotron emission from relativistic electrons in a power-law distribution ($dn_e/d\gamma_e\propto \gamma_e^{-p}$), with a constant power index, $p=2.17$, between 10 to 1000 days with no sign of spectral evolution \citep{hallinan_radio_2017,troja_x-ray_2017,lyman_optical_2018,margutti_binary_2018,troja_thousand_2020,makhathini_panchromatic_2021,balasubramanian_continued_2021,balasubramanian_gw170817_2022}. The spectrum follows $F_\nu\propto \nu^{(1-p)/2}$ at all times, indicating that the radio emission surpasses both the peak frequency, $\nu_m$, and the self-absorption frequency, $\nu_a$, while the X-ray emission lie below the cooling frequency, $\nu_c$. Notably, it means that the frequencies $\nu_m$ and $\nu_a$ are below $\sim1$GHz \citep{dobie_turnover_2018,balasubramanian_continued_2021,balasubramanian_gw170817_2022}, and $h\nu_c$ is above $\sim10$keV \citep{troja_thousand_2020,troja_accurate_2022} between 10 to 1000 days. These bounds constrain the source velocity, the medium density, and the fraction of energy held by magnetic fields.

As mentioned in \citet{linial_cooling_2019}, relativistic electrons undergo energy loss through two distinct mechanisms: (i) synchrotron emission and (ii) up-scattering of photons by inverse-Compton processes. The latter process relies on the presence of low-energy photons. During the initial 10 days, the KN serves as such a photon source, necessitating consideration of inverse-Compton cooling. In general, the low-energy tail of the synchrotron emission also serves as a source of seed photons for synchrotron self-Compton (SSC) processes. However, the potential upper limit for the cooling frequency derived from SSC is significantly higher. This is because the cooling frequency is inversely proportional to the square of the photon energy density, which is dominated by the KN.

A superluminal centroid motion was observed in radio, and together with the HST localization of the early thermal component, it provides 4 different center of light localizations between $8$ days to $230$ days   \citep{mooley_superluminal_2018,ghirlanda_compact_2019,mooley_optical_2022}, implying an off-axis synchrotron emission of a relativistic outflow. The presence of plasma expanding at $\gamma>10$ (which is assumed by structured jet models) cannot be directly inferred from observations, which may be accounted for by a $\gamma\gtrsim5$ outflow \citep{mooley_optical_2022}. 
The synchrotron break frequencies for an ultra-relativistic spherical blast wave were derived by \citet{granot_shape_2002}. 
\citet{linial_cooling_2019} estimated the effect of synchrotron and inverse-Compton cooling in the case of GW170817 while considering on-axis emission. They concluded that 9 days post-merger, the Lorentz factor, $\gamma$, must exceed 2.1.
GW170817 occurred in an elliptical, low-density galaxy, NGC 4993 \citep{abbott_multi-messenger_2017}; future merger events may occur in non-elliptical galaxies, with much higher ISM densities, $n\sim1$cm$^{-3}$. In such a case, the break frequencies are expected to be observed in radio and X-ray bands. Thus, accurately describing the synchrotron break frequencies would provide stringent constraints over the outflow velocity, observing angle, ISM density, and fractions of energy held by relativistic electrons and magnetic fields in the shocked plasma. 
Naturally, most of the mergers will be observed off-axis, and consequently, the emission is expected to be dominated by $\gamma<10$ shocked plasma due to relativistic beaming.

Off-axis relativistic jets with an observed afterglow in which a gamma-ray burst is not detected are known as orphan afterglows \citep{granot_off-axis_2002}. These events are among the promising electromagnetic counterparts to compact objects merger \citep{metzger_what_2012}. Recently, such an afterglow was observed to have a Lorentz factor of $\gamma<10$ \citep{perley_at2019pim_2024}. Additionally, there are indications that tidal disruption events can also generate mildly-to-highly relativistic outflows \citep{burrows_relativistic_2011,saxton_xmmsl1_2017,alexander_radio_2020,andreoni_very_2022,beniamini_swift_2023,rhodes_day-time-scale_2023}. Providing simple analytic expressions for the break frequencies for these types of outflows is essential for accurately determining their properties.

We derive the synchrotron break frequencies for a uniform volume element of shocked plasma propagating at mildly to highly relativistic velocities, considering both on-axis and off-axis observers. In this description, we assume that the velocity direction of the shocked plasma is perpendicular to the shock front (the normal), or equivalently, $v_\perp\gg v_\parallel$, where $v_\perp/v_\parallel$ is the lab frame shocked plasma velocity component perpendicular/parallel to the shock front. It applies to various relativistic astrophysical phenomena, including: on-axis emission from a relativistic jet, off-axis emission from a structured jet for as long as the structure propagates radially (\citet{govreen-segal_structure_2023} showed that a power-law jet structure with $dE/d\theta\propto \theta^{\gtrsim-3}$, keeps its initial shape for $\gamma\geq 3$, indicating that the evolution of each angular bin can be approximated by Blandford-McKee evolution \citep{blandford_fluid_1976} such that the lateral spreading is negligible), off-axis emission from a relativistic jet as the Lorentz factor reaches $\gamma\sim1/\theta_v$ (where $\theta_v$ is the viewing angle) and for a shocked plasma driven by ejecta that propagates into the ISM, such that the shock propagation can be approximated by a section of a spherical outflow \citep{sadeh_non-thermal_2024-1}.
For the self-absorption frequency, a specific geometrical structure is assumed. We further consider a power-law energy distribution for electrons, given by $dn_e/d\gamma_e\propto \gamma_e^{-p}$. Here, $\gamma_e$ represents the electron Lorentz factor within the plasma's rest frame, while $p$ is within the range of $2\leq p\leq2.5$. Simultaneously, fractions $\varepsilon_e$ and $\varepsilon_B$ correspond to the post-shock internal energy density proportions attributed to non-thermal electrons and magnetic fields, respectively. 
This phenomenological description, capturing the post-shock plasma conditions, finds support across a diverse spectrum of observations and plasma calculations encompassing both relativistic and non-relativistic shocks \citep[see][]{blandford_particle_1987,keshet_energy_2005,waxman_gamma-ray_2006,keshet_analytical_2006,bykov_fundamentals_2011,sironi_maximum_2013,pohl_pic_2020,ligorini_mildly_2021,kobzar_electron_2021}. 

The paper is organized as follows. In \S~\ref{sec:bounds}, we derive the analytic expressions for the synchrotron break frequencies. We compare our results to accurate analytic calculations and earlier estimations in \S~\ref{sec:earlier}.
Finally, in \S~\ref{sec:GW170817}, we discuss the implications of our analysis to GW170817 synchrotron observations, and our conclusions are summarized in \S~\ref{sec:conclusions}.

\section{Analytic expressions}
\label{sec:bounds}
\begin{figure}
\includegraphics[width=9cm]{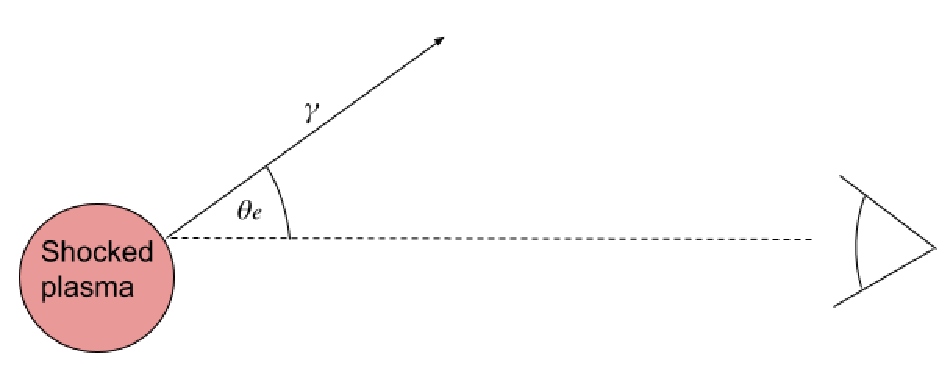}
    \caption{A schematic illustration of the conditions described in \ref{sec:bounds}.}
\label{fig:blob}
\end{figure}

We assume that the flux is primarily produced by a synchrotron-emitting blob consisting of uniformly shocked plasma with a Lorentz factor $\gamma$, moving through a homogeneous ISM with a number density $n$. Additionally, we assume that the dynamical evolution of the volume element is negligible while it remains the dominant source of the observed synchrotron emission. A spherical coordinate system is used, where the blob is located at a distance $R$ from the origin. The observer is positioned such that the angle between the blob's velocity and the line of sight is $\theta_e$ (see Fig. \ref{fig:blob}). 
We employ a phenomenological approach to describe synchrotron emission: the fraction of energy carried by electrons is denoted by $\varepsilon_e$, and the fraction of energy carried by magnetic fields is denoted by $\varepsilon_B$. We assume a power-law electron distribution within the shocked layers, $dn_e/d\gamma_e\propto \gamma_e^{-p}$.
The characteristic plasma frame frequency of the synchrotron emission of electrons with Lorenz factor $\gamma_e$ and isotropic velocity distribution is \citep{rybicki_radiative_1979}
 \begin{equation}
     \nu_s\approx\frac{1}{3}\cdot\frac{3}{4\pi}\langle\sin\alpha\rangle\gamma_e^2\frac{q_eB}{m_ec}\approx \gamma_e^2\frac{q_e\sqrt{u_B}}{3m_ec},
 \end{equation}
where the $1/3$ factor is due to considering the peak emission in the spectrum and $u_B$ is the (plasma frame) magnetic field energy density. Approximating the post shock energy density, $2nm_p v^2$ (where $v$ is the plasma velocity) for non-relativistic and $4\gamma^2nm_p c^2$ for highly relativistic shocks, by $4\gamma(\gamma-1)nm_p c^2$, we have $u_B=4\varepsilon_B\gamma(\gamma-1)n m_p c^2$.
\begin{equation}
\nu_s=0.7\gamma_e^2\frac{q_e\sqrt{\gamma(\gamma-1)\varepsilon_Bnm_p}}{m_e}.
\end{equation}
In the lab frame, we have\footnote{The first order corrections in $v_\parallel/v_\perp$ are $\theta_e\rightarrow\theta_e\pm\frac{v_\parallel}{v_\perp}$, $\beta\rightarrow\beta$, $\gamma\rightarrow\gamma$.}
\begin{equation}
\label{eq:nu}
    \nu=\frac{\nu_s}{\gamma(1-\beta\cos(\theta_e))}=0.7\gamma_e^2\sqrt{\frac{\gamma-1}{\gamma}}\cdot\frac{q_e\sqrt{\varepsilon_Bnm_p}}{m_e(1-\beta\cos(\theta_e))}.
\end{equation}
In this estimation, we assume the emitting region is relatively uniform and that the shocked material right behind the shock is dominating the synchrotron emission; these assumptions were validated for ultra-relativistic flows \citep{waxman_gamma-rayburst_1997,waxman_angular_1997,granot_shape_2002} and mildly-relativistic flows \citep{sadeh_non-thermal_2023,sadeh_non-thermal_2024-1,sadeh_late-time_2024}. In realistic scenarios, there are various contributions to the flux from different angles and shocked plasma velocities. However, three points should be noted here: first, if the break frequencies are not observed, as in the case of GW170817, the bounds provided by this analysis would be consistent with all of the various optional $\theta_e$ and $\gamma$. Second, if indeed there are competing contributions from two separate regions of considerably different $\theta_e$ and $\gamma$, the break frequencies would be significantly smeared, and this analysis can be used to reconstruct such a smear in the spectrum. Third, in case the break frequencies are observed with distinct values, it is reasonable to assume most of the contribution arrives from tightly distributed values of $\theta_e$ and $\gamma$.   
\subsection{Peak frequency}
For a power-law distribution of electrons $dn_e/d\gamma_e\propto \gamma_e^{-p}$ (between $\gamma_\text{min}$ and $\gamma_\text{max}$) 
$\gamma_{\rm min}$ is obtained by requiring the average energy per electron to equal a fraction $\varepsilon_e$ of the post-shock internal energy per particle, $(\gamma-1)m_pc^2$.
the minimal energy electrons Lorentz factor is \citep{sadeh_non-thermal_2023}
\begin{equation}
  \gamma_{\rm min}=\frac{l_p(\gamma-1)}{p-1}\frac{\varepsilon_em_p}{m_e},
\end{equation}
where
\begin{equation}\label{eq:lp}
  l_p=\frac{p-2}{1-(\gamma_\text{max}/\gamma_\text{min})^{2-p}},\quad
  l_p\xrightarrow[p \to 2]{}\frac{1}{\ln(\gamma_\text{max}/\gamma_\text{min})}.
\end{equation}
The observed frequency from these electrons is (from Eq. (\ref{eq:nu}))
\begin{equation}
\begin{aligned}
  \nu_m&=0.7\left(\frac{l_p(\gamma-1)}{p-1}\frac{\varepsilon_em_p}{m_e}\right)^2\times\\&\sqrt{\frac{\gamma-1}{\gamma}}\cdot\frac{q_e\sqrt{\varepsilon_Bnm_p}}{m_e(1-\beta\cos(\theta_e))}.
  \end{aligned}
\end{equation}
For $p=2.17$
\begin{equation}
\label{eq:num}
    \nu_m=1.5\varepsilon^2_{e,-1}\varepsilon^\frac{1}{2}_{B,-2}n^{\frac{1}{2}}_{-3}\frac{\gamma^{-\frac{1}{2}}(\gamma-1)^{\frac{5}{2}}}{(1-\beta\cos(\theta_e))}\text{MHz},
\end{equation}
where the prefactor varies up to a factor of $\sim4$ for $2<p<2.5$.
\subsection{Self-absorption frequency}
Following \citet{sadeh_non-thermal_2024-1}
we estimate the self-absorption frequency at time $t$ as the frequency for which $\tau_\nu=\alpha_\nu\Delta_\tau=1$, where $\alpha_\nu$ and $\Delta_\tau$ are the typical absorption coefficient and the typical path length traversed by photons through the shocked plasma, when the forward shock reached the radius $R$, dominating the emission of radiation observed at time $t$. In this analysis, we approximate the geometry of the shocked plasma as a narrow conical section of a spherical blast wave with an angle $\theta_e$ between the cone symmetry axis and the line of sight.
We estimate the thickness $\Delta_R$ of the emitting layer by conservation of particle number as
\begin{equation}
    \Delta_R\approx \frac{R}{12\gamma^2},
\end{equation}
and
\begin{equation}
    \Delta_\tau\approx\frac{\Delta_R}{\cos\theta_e}\approx\frac{\beta ct}{(12\gamma^2\cos(\theta_e))(1-\beta\cos(\theta_e))}.
\end{equation}
The $1/12$ factor is accurate to $\sim30\%$ for $\gamma>2$ shocked plasma. It is valid for both a forward-reverse shock structure, and for a relativistic blast wave propagating radially (without an ejecta propagation behind it). 
To derive the self-absorption frequency in the observer frame (primed, $'$, quantities are in the shocked plasma frame), we first approximate the absorption coefficient, which is given by \citep{rybicki_radiative_1979}
\begin{equation}
    \alpha_\nu = \frac{\nu'}{\nu}\alpha_\nu'\approx \gamma^{\frac{2-p}{4}}\left(1-\beta\cos(\theta_e)\right)^{-\frac{p+2}{2}}(\gamma-1)^{\frac{5p-2}{4}}f_a(p)\nu^{-\frac{p+4}{2}},
\end{equation}
where 
\begin{equation}
\begin{aligned}
       f_a(p)&=(p-1)\left(\frac{p-2}{p-1}\frac{\frac{\varepsilon_e m_p}{m_e} }{1-\left(\gamma_\text{max}/\gamma_\text{min}\right)^{2-p}}\right)^{p-1}\times\\
       &4n \left(\frac{2\pi m_e c}{3q_e}\right)^{-\frac{p}{2}} \frac{(32\pi\varepsilon_B nm_p c^2)^{\frac{p+2}{4}}\sqrt{3}q_e^3}{8\pi m_e^2c^2}\times\\
       &\frac{\sqrt{\pi}}{2}\frac{\Gamma\left(\frac{p+6}{4}\right)}{\Gamma\left(\frac{p+8}{4}\right)}\Gamma\left(\frac{3p+22}{12}\right)\Gamma\left(\frac{3p+2}{12}\right). 
\end{aligned}
\end{equation}
The optical depth at (observed) frequency $\nu$ is given by
\begin{equation}
\begin{aligned}
    \tau_\nu&\approx \alpha_\nu\Delta_\tau\approx \\&\frac{\beta ct \gamma^{\frac{2-p}{4}}\left(1-\beta\cos(\theta_e)\right)^{-\frac{p+2}{2}}(\gamma-1)^{\frac{5p-2}{4}}f_a(p)\nu^{-\frac{p+4}{2}}}{(12\gamma^2\cos(\theta_e))(1-\beta\cos(\theta_e))}.
    \end{aligned}
\end{equation}
The self-absorption frequency, $\nu_{a}$ defined by $\tau_\nu(\nu=\nu_{a})=1$, is finally given by
\begin{equation}
    \nu_a = \left(\frac{\beta ct \gamma^{\frac{2-p}{4}}(\gamma-1)^{\frac{5p-2}{4}}f_a(p)}{12\gamma^2\cos(\theta_e)(1-\beta\cos(\theta_e))^{\frac{p+4}{2}}} \right)^{\frac{2}{p+4}}.
\end{equation}

\subsection{Synchrotron Cooling frequency}
$\gamma_c^\text{syn}$ is obtained by requiring the (plasma frame) synchrotron loss time \citep{sadeh_non-thermal_2024-1}, $m_ec^2/(\gamma_e(4/3) \sigma_T cu_B)$, to be equal to the time measured at the plasma frame, $t/\gamma/\left(1-\beta\cos(\theta_e)\right)$.
\begin{equation}
    \gamma_c^\text{syn}=
  \frac{\left(1-\beta\cos(\theta_e)\right)m_e}{4(\gamma-1)\varepsilon_B\left(\frac{4}{3}\right)\sigma_T n m_p ct}.
\end{equation}
Notice that this derivation assumes that $\int \beta dt\approx \beta\cdot t$, an order unity factor is ignored, and the formula has to be tested for the specific cases we consider. It is shown to be accurate for a case of shocked plasma driven by ejecta propagating radially in \citet{sadeh_non-thermal_2024-1} and for an ultra-relativistic blast wave propagating radially in \S~\ref{sec:earlier}.
The observed frequency from these electrons is
\begin{equation}
  \begin{aligned}
h\nu_c^\text{syn}&=0.7\left(\frac{\left(1-\beta\cos(\theta_e)\right)m_e}{4(\gamma-1)\varepsilon_B\left(\frac{4}{3}\right)\sigma_T n m_p ct}\right)^2\times\\&\sqrt{\frac{\gamma-1}{\gamma}}\cdot\frac{hq_e\sqrt{\varepsilon_Bnm_p}}{m_e(1-\beta\cos(\theta_e))}.
  \end{aligned}
\end{equation}
Which can be written as
\begin{equation}
\begin{aligned}
    \label{eq:nuc}
h\nu_c^\text{syn}=9.1t_{155}^{-2}\varepsilon^{-\frac{3}{2}}_{B,-2}n^{-\frac{3}{2}}_{-3}\frac{1-\beta\cos(\theta_e)}{\gamma^{\frac{1}{2}}(\gamma-1)^{\frac{3}{2}}}\text{keV},
\end{aligned}
\end{equation}
where $t_{155}=t/155$days.
\subsection{Inverse Compton cooling}
$\gamma_c^\text{IC}$ is obtained by requiring the (plasma frame) inverse-Compton loss time, $m_ec^2/(\gamma_e(4/3) \sigma_T cu_\gamma)$, to be equal to the time measured at the plasma frame, $t/\gamma/\left(1-\beta\cos(\theta_e)\right)$. 
\begin{equation}
    \gamma_c^\text{IC}=
  \frac{\gamma\left(1-\beta\cos(\theta_e)\right)m_e c}{u_\gamma\left(\frac{4}{3}\right)\sigma_T t}=\frac{\gamma^3\beta^2 3\pi m_e c^4t}{\sigma_T L_\text{bol}\left(1-\beta\cos(\theta_e)\right)},
\end{equation}
where the photon energy density, $u_\gamma$, is given by \citep{linial_cooling_2019}
\begin{equation}
    u_\gamma=\frac{L_\text{bol}}{4\pi R^2 c}\cdot\frac{1}{\gamma^2}=u_\gamma=\frac{L_\text{bol}}{4\pi c^3t^2}\cdot\frac{(1-\beta\cos(\theta_e))^2}{\gamma^2\beta^2}.
\end{equation}
The observed frequency from these electrons is
\begin{equation}
\begin{aligned}
h\nu_c^\text{IC}&=0.7\left(\frac{\gamma^3\beta^2 3\pi m_e c^4t}{\sigma_T L_\text{bol}\left(1-\beta\cos(\theta_e)\right)}\right)^2\times\\&\sqrt{\frac{\gamma-1}{\gamma}}\cdot\frac{hq_e\sqrt{\varepsilon_Bnm_p}}{m_e(1-\beta\cos(\theta_e))}.
\end{aligned}
\end{equation}
Which can be written as
\begin{equation}
\begin{aligned}
\label{eq:IC}
h\nu_c^\text{IC}=5.1L_{\text{bol},40}^{-2}t_{10}^{2}\varepsilon^{\frac{1}{2}}_{B,-2}n^{\frac{1}{2}}_{-3}\frac{\gamma^{\frac{11}{2}}\beta^4(\gamma-1)^{\frac{1}{2}}}{(1-\beta\cos(\theta_e))^3}\text{eV},
\end{aligned}
\end{equation}
where $L_{\text{bol},40}=L_{\text{bol}}/10^{40}$erg/s. 

\section{Comparison to earlier results}
\label{sec:earlier}
\subsection{Full calculation of spherical ultra-relativistic blast wave}
\begin{figure}
    \centering
    \includegraphics[width=\columnwidth]{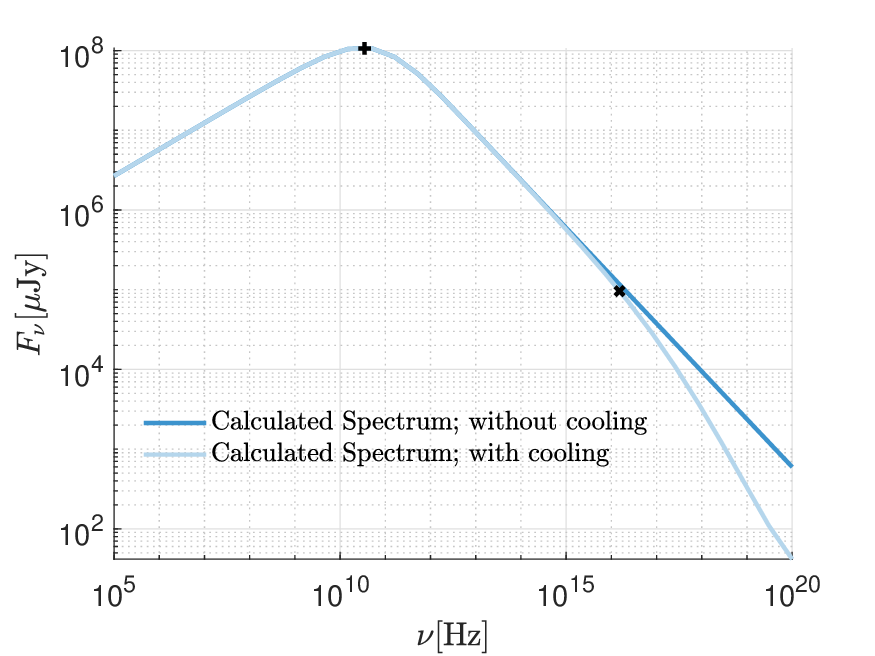}
    \caption{Calculated synchrotron spectrum following \citet{gruzinov_gamma-ray_1999} for the following parameters $\{E,n,t,\varepsilon_e,\varepsilon_B,p\}=\{10^{55}\text{erg},1\text{cm}^{-3},2\text{days},5\times10^{-2},10^{-4},2.2\}$ with and without synchrotron cooling. The analytic $\nu_m$ and $\nu_c$, Eqs. (\ref{eq:num}) and (\ref{eq:nuc}), are shown by '+' and 'x' signs respectively.}
    \label{fig:frequency_check}
\end{figure}

In \citet{gruzinov_gamma-ray_1999}, an accurate analytic calculation is provided for the synchrotron light curve and spectrum of a spherical ultra-relativistic strong explosion \citep{blandford_fluid_1976}. In Fig. \ref{fig:frequency_check}, we provide the full accurate spectrum calculated for a given set of parameters $\{E,n,t,\varepsilon_e,\varepsilon_B,p\}=\{10^{55}\text{erg},1\text{cm}^{-3},2\text{days},5\times10^{-2},10^{-4},2.2\}$, following \citet{gruzinov_gamma-ray_1999} with and without synchrotron cooling. The proper relation between the observed time and the shocked plasma Lorentz factor, in this case, is $t\approx R/4\gamma^2c$, where $R$ is the shock radius \citep{sari_spectra_1998,sari_observed_1998}. Since the explosion is ultra-relativistic, $\gamma\gg 1$, we use $\cos\theta_e\approx1$. Thus the relation between the emitting plasma Lorentz factor and the observed time is \citep{blandford_fluid_1976}
\begin{equation}
\label{eq:spherical}
    \gamma = \left(\frac{17}{1024\pi}\cdot\frac{E}{n m_p c^5}\cdot\frac{1}{t^3}\right)^{\frac{1}{8}},
\end{equation}
where we assumed here that most of the emission arrives from the shocked plasma right behind the shock. Our analytic estimations for $\nu_m$ and $\nu_c$ are added to Fig. \ref{fig:frequency_check} with good agreement.
\subsection{Previous analytical estimations}
\begin{figure}
    \centering
    \includegraphics[width=\columnwidth]{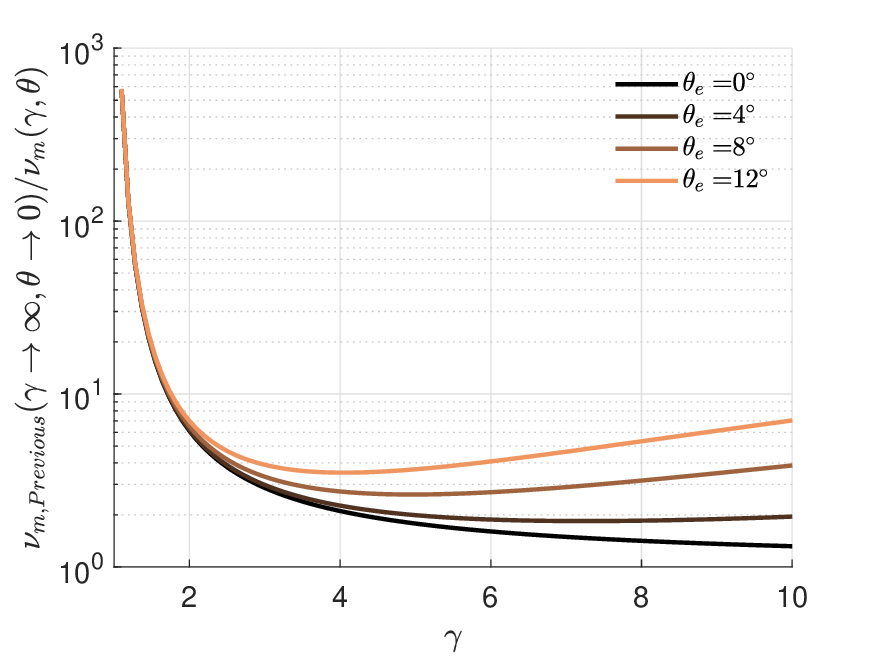}
    \caption{The ratio between the analytical estimation from \citet{granot_shape_2002} for $\nu_m$, which is valid in the limit of $\theta=0,\gamma\rightarrow\infty$, to our extension given in Eq. (\ref{eq:num}) for various Lorentz factors and viewing angles. We use Eq. (\ref{eq:spherical}) to adjust to the spherical ultra-relativistic strong explosion.}
    \label{fig:nu_m_com}
\end{figure}

In \citet{granot_shape_2002}, analytic estimation for $\nu_m$ was presented for the case of a spherical ultra-relativistic strong explosion \citep{blandford_fluid_1976} that is later used in the context of GW170817 by \citet{gill_numerical_2019}. In Fig. \ref{fig:nu_m_com}, we show the ratio between this estimation to our extension from Eq. (\ref{eq:num}) using Eq. (\ref{eq:spherical}) for various Lorentz factors and off-axis observers.

\begin{figure}
    \centering
    \includegraphics[width=\columnwidth]{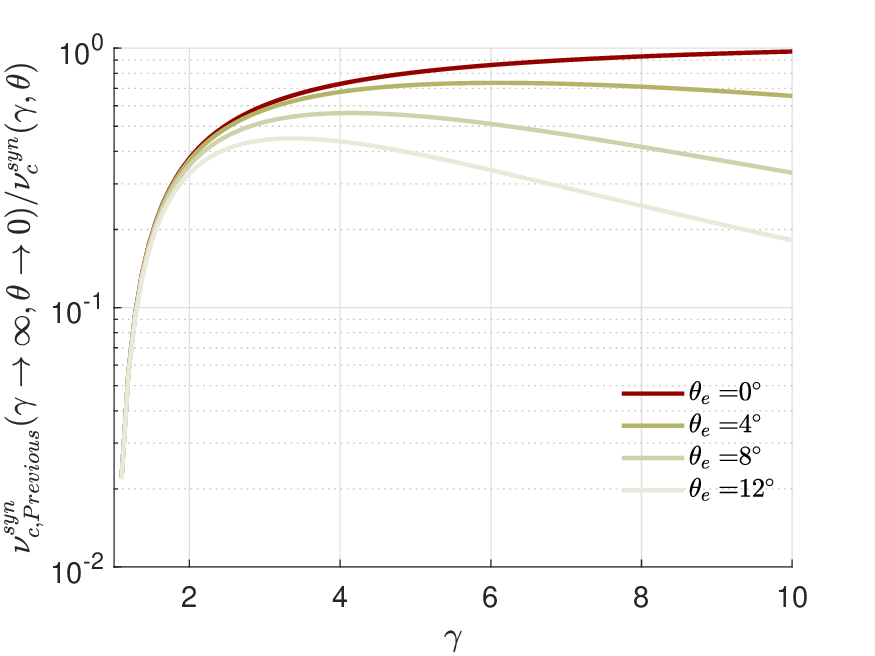}
    \caption{The ratio between the analytical estimation from \citet{linial_cooling_2019} for $\nu_c^\text{syn}$, which is valid in the limit of $\theta=0,\gamma\rightarrow\infty$, to our extension given in Eq. (\ref{eq:nuc}) for various Lorentz factors and viewing angles.}
    \label{fig:nu_c_com}
\end{figure}

\begin{figure}
    \centering
    \includegraphics[width=\columnwidth]{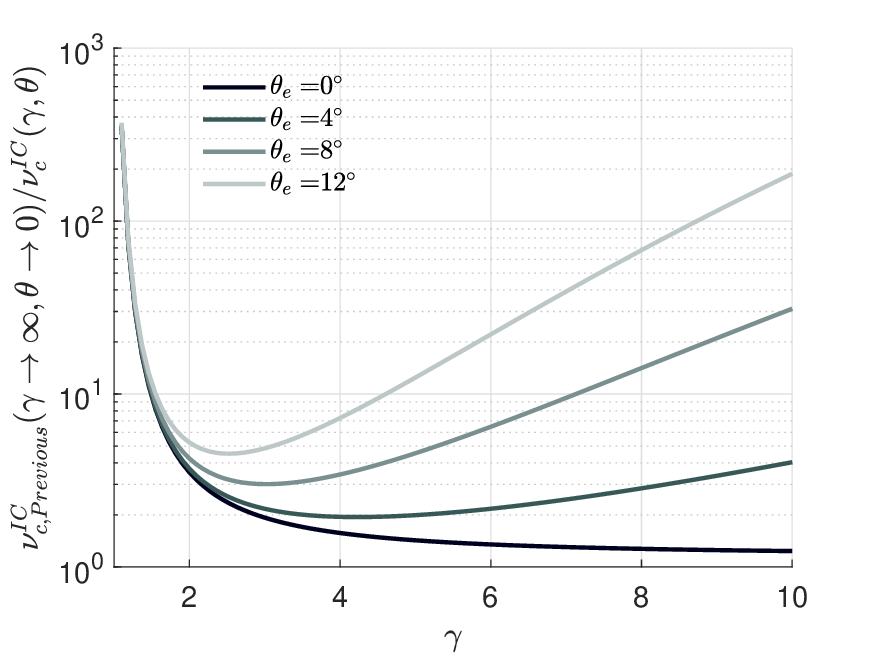}
    \caption{The ratio between the analytical estimation from \citet{linial_cooling_2019} for $\nu_c^\text{IC}$, which is valid in the limit of $\theta=0,\gamma\rightarrow\infty$, to our extension given in Eq. \ref{eq:IC} for various Lorentz factors and viewing angles.}
    \label{fig:nu_IC_com}
\end{figure}

In \citet{linial_cooling_2019}, analytic estimations for both $\nu_c^\text{syn}$ and $\nu_c^\text{IC}$ are presented for the case of ultra-relativistic outflow emitting on-axis, later used in the context of GW170817. In Figs. \ref{fig:nu_c_com} - \ref{fig:nu_IC_com}, we show the ratio between those estimations to our extensions in Eqs. (\ref{eq:nuc}) and (\ref{eq:IC}) for various Lorentz factors and off-axis observers.

\section{GW170817 Implications}
\label{sec:GW170817}
The radio to X-ray observations of GW170817 between $10-1000$ days are consistent with a spectrum of $F_\nu\propto \nu^{(1-p)/2}$, with $p=2.17$, implying $\nu_m,\nu_a<2.5$GHz \citep{dobie_turnover_2018,balasubramanian_continued_2021,balasubramanian_gw170817_2022} and $h\nu^{\text{syn}}_c,h\nu^{\text{IC}}_c>10$keV \citep{troja_thousand_2020,troja_accurate_2022}. The observed superluminal centroid motion between $70$ to $230$ days indicates $\theta_e>5^\circ$ and $\gamma\geq4$ \citep{mooley_superluminal_2018,mooley_optical_2022}. Emission from shocked plasma with Lorentz factor $\gamma$ would be dominant only for angles $\theta_e<\sin^{-1}\left(\frac{1}{\gamma}\right)$ due to relativistic beaming. Considering all of the above, we provide bounds for the shocked plasma parameter space. There is no constraint from $\nu_a$ due to the low ISM density.
\subsection{Synchrotron break frequencies constraints}
\subsubsection{$\nu^\text{syn}_c$}
\begin{figure}
    \centering
    \includegraphics[width=\columnwidth]{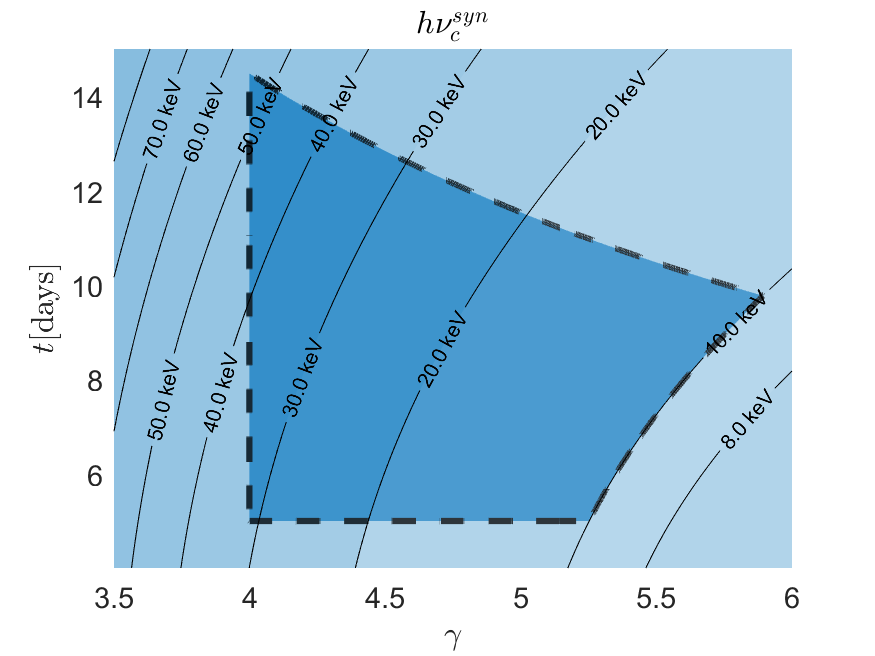}
    \caption{In dark blue: The allowed parameter space for $h\nu^\text{syn}_c>10$keV, $\theta_e>5^\circ$, $\gamma\geq4$, and $\theta_e<\sin^{-1}\left(\frac{1}{\gamma}\right)$. The black lines represent equal $h\nu^\text{syn}_c$ curves. We used $\varepsilon_{B,-2}n_{-3}=0.01$, $3.5<\gamma<6$, and $t_{155}=1$}
    \label{fig:nu_c}
\end{figure}

From Eq. \ref{eq:nuc} we find
\begin{equation}
    9.1t_{155}^{-2}\varepsilon^{-\frac{3}{2}}_{B,-2}n^{-\frac{3}{2}}_{-3}\frac{1-\beta\cos(\theta_e)}{\gamma^{\frac{1}{2}}(\gamma-1)^{\frac{3}{2}}}>10.
\end{equation}
At 155 days (the peak time), we have
\begin{equation}
\label{eq:nu_c_b}
    \varepsilon^{-\frac{3}{2}}_{B,-2}n^{-\frac{3}{2}}_{-3}\frac{1-\beta\cos(\theta_e)}{\gamma^{\frac{1}{2}}(\gamma-1)^{\frac{3}{2}}}>1.1.
\end{equation}
Considering the abovementioned bounds, Eq. \ref{eq:nu_c_b} has a solution only for $\varepsilon_{B,-2}n_{-3}\leq0.03$. In Fig. \ref{fig:nu_c} we show an example for the possible parameter space for $\varepsilon_{B,-2}n_{-3}=0.01$, $3.5<\gamma<6$, $t_{155}=1$ and $4^\circ<\theta_e<15^\circ$.

\subsubsection{$\nu^\text{IC}_c$}
\begin{figure}
    \centering
    \includegraphics[width=\columnwidth]{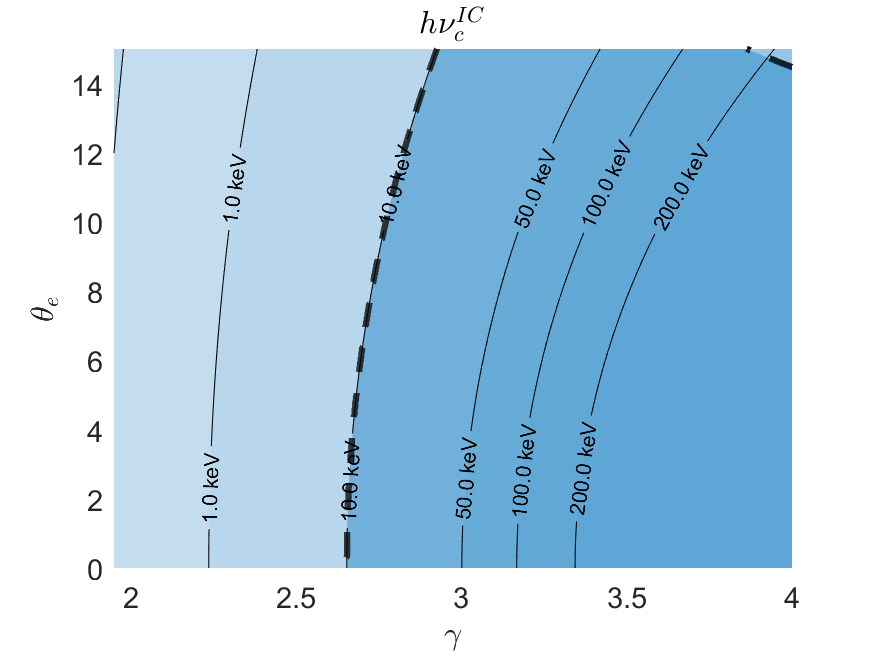}
    \caption{In dark blue: The allowed parameter space for $h\nu^\text{IC}_c>10$keV. The black lines represent equal $h\nu^\text{IC}_c$ curves. We used $\varepsilon_{B,-2}n_{-3}=0.03$ and $t=9$.}
    \label{fig:nu_IC}
\end{figure}

From Eq. \ref{eq:IC} we find
\begin{equation}
5.1L_{\text{bol},40}^{-2}t_{10}^{2}\varepsilon^{\frac{1}{2}}_{B,-2}n^{\frac{1}{2}}_{-3}\frac{\gamma^{\frac{11}{2}}\beta^4(\gamma-1)^{\frac{1}{2}}}{(1-\beta\cos(\theta_e))^3}>10000.
\end{equation}
Since we know that at $9$days The bolometric luminosity from the KN is $L_\text{bol}=6\times10^{40}$ \citep{linial_cooling_2019} we can write
\begin{equation}
\varepsilon^{\frac{1}{2}}_{B,-2}n^{\frac{1}{2}}_{-3}\frac{\gamma^{\frac{11}{2}}\beta^4(\gamma-1)^{\frac{1}{2}}}{(1-\beta\cos(\theta_e))^3}>87757,
\end{equation}
In Fig. \ref{fig:nu_IC} we show the possible parameter space for $\varepsilon_{B,-2}n_{-3}=0.03$, $1<\gamma<10$, $t=9$ days and $0^\circ<\theta_e<10^\circ$. We find a lower bound for the Lorentz factor of $\gamma>2.6$.

\subsubsection{$\nu_m$}
\begin{figure}
    \centering
    \includegraphics[width=\columnwidth]{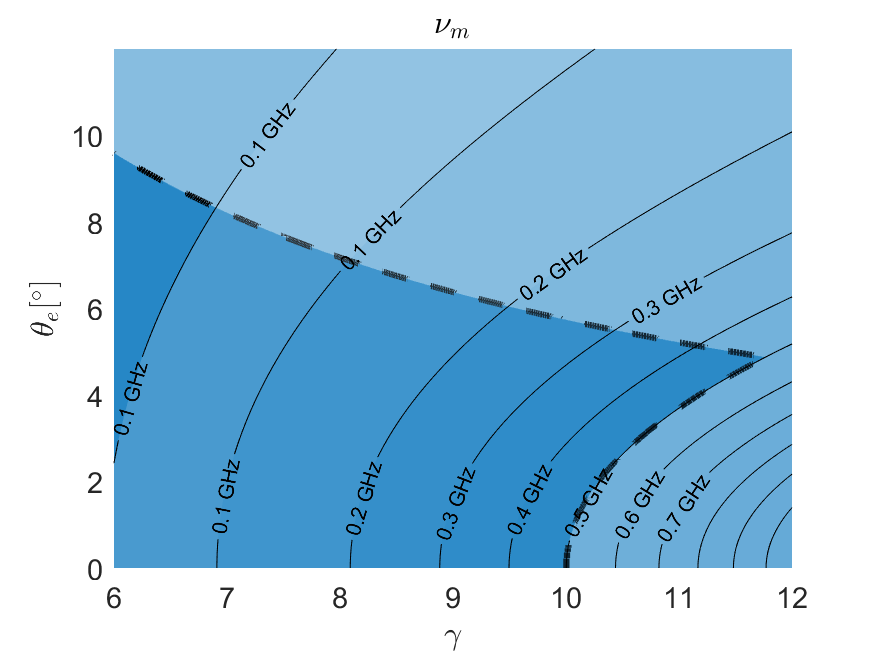}
    \caption{In dark blue: The allowed parameter space for $\nu_m>0.5$GHz, and $\theta_e<\sin^{-1}\left(\frac{1}{\gamma}\right)$. The black lines represent equal $\nu_m$ curves. We used $p=2.17$ and $\varepsilon^2_{e,-1}\varepsilon^\frac{1}{2}_{B,-2}n^{\frac{1}{2}}_{-3}=0.07$.}
    \label{fig:nu_m}
\end{figure}

The actual electron distribution in realistic conditions is not a pure power-law; the electrons are expected to be in a thermal population with a high energy power-law tail. As such, a proper bound for GW170817 $\nu_m$ would be $\nu_m\ll2.5$GHz. Using a bound of $\nu_m<0.5$GHz together with Eq. (\ref{eq:num}) we find
\begin{equation}
\varepsilon^2_{e,-1}\varepsilon^\frac{1}{2}_{B,-2}n^{\frac{1}{2}}_{-3}\frac{\gamma^{-\frac{1}{2}}(\gamma-1)^{\frac{5}{2}}}{(1-\beta\cos(\theta_e))}<731.
\end{equation}
In Fig. \ref{fig:nu_m}, we show the possible parameter space for $p=2.17$, $\varepsilon^2_{e,-1}\varepsilon^\frac{1}{2}_{B,-2}n^{\frac{1}{2}}_{-3}=0.07$, $6<\gamma<12$ and $0^\circ<\theta_e<12^\circ$. The values of $\varepsilon^2_{e,-1}\varepsilon^\frac{1}{2}_{B,-2}n^{\frac{1}{2}}_{-3}=0.07$ are considerably small so we find it reasonable to put an upper bound over Lorentz factor of $\gamma<12$ for the emitting shocked plasma for $t>16$ days (first radio observations).

\subsection{Previous fits to GW170817}
Some of the previous fits to the synchrotron observations of GW170817 provide values in tension with the abovementioned bounds. For example, in many semi-analytical calculations, the best fits found indicate $\varepsilon_{B,-2}n_{-3}=0.1-1$, suggesting that it is essential that the bounds from the spectrum analysis should be incorporated into the fitting pipeline. Furthermore, it is important to notice that the potential jet structure that fits the rising non-thermal light curve of GW170817 is consistent with the lower bounds over the shocked plasma Lorentz factor, $\gamma>2.6$. These bounds are consistent with the suggested alternative in \citet{sadeh_non-thermal_2024}, a conical, radially-stratified outflow.
\section{Conclusions}
\label{sec:conclusions}
In this work, we derived analytical expressions for the synchrotron break frequencies, namely $\nu_m$, $\nu_a$, $\nu_c^\text{syn}$ and $\nu_c^\text{IC}$ from shocked plasma that are valid for both mildly-relativistic and ultra-relativistic outflows and for both on- and off-axis observers, extending the validity of previous analytic estimations which were valid for on-axis emission and ultra-relativistic velocities (see Figs. 
\ref{fig:nu_m_com}-\ref{fig:nu_IC_com} in \S~\ref{sec:earlier}). Our analytic formulae for $\nu_m$ and $\nu_c^\text{syn}$ were tested successfully for the fully accurate analytic solution in the case of an ultra-relativistic spherical blast wave (see Fig. \ref{fig:frequency_check}). Our analytic formulae for $\nu_c^\text{IC}$ is based on the same calculation as $\nu_c^\text{syn}$, and our analytic formula for $\nu_a$ was successfully tested for specific cases of mildly relativistic ejecta in \citet{sadeh_non-thermal_2024-1}.

These results will be useful for accurate parameter estimation ($n$, $\gamma$, $\theta_e$, $\varepsilon_B$, and $\varepsilon_e$) from future merger events and jetted TDEs since most of the them will be observed off-axis, and the synchrotron emission is expected to be dominated by $\gamma<10$ shocked plasma due to the relativistic beaming.
In typical ISM densities, $(n\sim1\text{cm}^{-3})$, the break frequencies are expected to be observed in radio and X-ray. Consequently, the analytic formulae derived in this paper would provide more significant constraints than those derived for GW170817.

Applying our derived analytic formulae for the case of GW170817 provided bounds over the ISM density, $n$, and the fraction of energy held by magnetic fields $\varepsilon_B$ such that $n\cdot\varepsilon_{B}\leq3\times10^{-7}\text{cm}^{-3}$. The Lorentz factor of the emitting shocked plasma is bounded by $\gamma>2.6$ at $t\sim10$ days and $\gamma<9.5$ at $t<230$ days. Some of the results trying to fit the synchrotron emission data of GW170817 are in tension with these bounds (see \S \ref{sec:GW170817}), while it is consistent with the solution suggested in \citet{sadeh_non-thermal_2024}.
For any robust attempt to fit future similar events, it's essential to take into account the synchrotron break frequencies. This consideration is crucial to ensure the model's validity, particularly in light of the non-trivial constraints discussed above.

The effect of the thermal population of electrons has been explored in \citet{margalit_thermal_2021,margalit_peak_2024}, suggesting it could have a significant impact the emission spectrum at frequencies of $\nu\gtrsim1$GHz in cases of mildly relativistic shocks propagating in ISM densities of $\gg1$cm$^{-3}$ with $\varepsilon_e\ll0.1$. Such conditions are primarily relevant to phenomena like "Fast Blue Optical Transients" \citep{ho_luminous_2022}. For instance, the late ($\lesssim1000$ days) non-thermal emission following GW170817 is consistent with emission from a power-law distribution of electrons due to the relatively low ISM density, $n\ll1$cm$^{-3}$, despite the shocked plasma being mildly relativistic.
Nevertheless, the thermal population does not influence the derivations obtained in this paper which are valid for the power-law population of electrons. Additionally, in this paper, we considered a constant ISM density, $n$, which may not apply to cases involving substantial mass loss via winds from the progenitor, which is beyond the scope of this paper. Such constant mass-loss wind is expected to produce $n\propto r^{-2}$ profiles altering some of the results discussed in this work. While the peak frequency and inverse Compton cooling frequency would remain unaffected, the self-absorption and synchrotron cooling frequencies would increase and decrease, respectively, due to the higher densities and, correspondingly, higher magnetic fields encountered during the expansion.

\section*{Acknowledgements}
We thank Eli Waxman and Andrei Gruzinov for their helpful contribution and insightful comments. 

\section*{Data Availability}
The data underlying this article will be shared following a reasonable request to the corresponding author.


\bibliography{references} 

\bibliographystyle{aasjournal}

\end{document}